\documentclass[prb,preprintnumbers,amsmath,amssymb,floatfix]{revtex4}

\usepackage{graphicx}
\usepackage{subfigure}
\usepackage{dcolumn}

\begin{document}

\title{Compound structure one-dimensional photonic crystal}
\author{Xiang-Yao Wu$^{a}$ \footnote{E-mail: wuxy2066@163.com},
 Ji Ma$^{a}$, Xiao-Jing Liu$^{a}$, Yu Liang$^{a}$, Zhi-Guo Wang$^{b}$ and Yun-Tuan Fang$^{c}$}
 \affiliation{a. Institute of Physics, Jilin Normal
University, Siping 136000 China\\
b. Pohl Institute of Solid State Physics, Tongji University,
Shanghai 20092, China\\
c. School of Computer Science and Telecommunication Engineering,
Jiangsu University, Zhenjiang 212013, China}

%%%%%%%
\begin{abstract}
In this paper, we have proposed a new compound structure
one-dimensional photonic crystal, which include series connection,
parallel connection and positive and negative feedback compound
structure photonic crystal. We have studied their transmission
characteristics and obtained some new results, which should be
help to design new type optical devices, such as optical
amplifier, photonic crystal laser and so on.

\vskip 5pt
PACS: 41.20.Jb, 42.70.Qs, 78.20.Ci\\
Keywords: photonic crystals; transmissivity; series connection;
parallel connection; positive and negative feedback
\end{abstract}
\maketitle

 \vskip 8pt
 {\bf 1. Introduction} \vskip 8pt
In 1987, E. Yablonovitch and S. John had pointed out that the
behavior of photons can be changed when propagating in the
material with periodical dielectric constant, and termed such
material Photonic Crystal [1, 2]. Photonic crystal important
characteristics are: Photon Band Gap, defect states, Light
Localization and so on. These characteristics make it able to
control photons, so it may be used to manufacture some high
performance devices which have completely new principles or can
not be manufactured before, such as high-efficiency semiconductor
lasers, right emitting diodes, wave guides, optical filters,
high-Q resonators, antennas, frequency-selective surface, optical
wave guides and sharp bends [3, 4], WDM-devices [5, 6], splitters
and combiners [7, 8]. optical limiters and amplifiers [9, 10]. The
research on photonic crystals will promote its application and
development on integrated photoelectron devices and optical
communication. To investigate the structure and characteristics of
band gap, there are many methods to analyze Photonic crystals
including the plane-wave expansion method [11], Green¡¯s function
method, finite-difference time-domain method [12-14] and transfer
matrix method [15-17].

In the paper, we have proposed a new compound structure
one-dimensional photonic crystal, which include series connection,
parallel connection and positive and negative feedback compound
structure photonic crystal. We have given their transmission
coefficients and transmissivities, studied their transmission
characteristics and obtained some new results: (1) The forbidden
band width of series connection photonic crystal becomes more
wider, and it is the union of corresponding forbidden band of part
photonic crystal, which is similar as the series connection ohm
law in circuit. (2) We can obtain the more wider forbidden band by
photonic crystal series connection. (3) With the number of series
connection photonic crystal increasing, the total width of
forbidden band increase. (4) The total forbidden band width of
parallel connection photonic crystal is adjusted by superposition
coefficient. (5) For the positive and negative feedback, the
transmissivity are larger than $1$, which can achieve light
amplification in the positive and negative feedback photonic
crystal. All these results should be help to design new type
optical devices, such as optical amplifier, photonic crystal laser
and so on.

 \vskip 8pt
 {\bf 2. Transfer matrix and transmissivity of one-dimensional
photonic crystal} \vskip 8pt

For one-dimensional conventional PCs, the calculations are
performed using the transfer matrix method [18], which is the most
effective technique to analyze the transmission properties of PCs.
For the medium layer $i$, the transfer matrices $M_i$ for $TE$
wave is given by [18]:
\begin{eqnarray}
M_{i}=\left(%
\begin{array}{cc}
 \cos\delta_{i} & -i\sin\delta_{i}/\eta_{i} \\
 -i\eta_{i}sin\delta_{i}
 & \cos\delta_{i}\\
\end{array}%
\right),
\end{eqnarray}
where $\delta_{i}=\frac{\omega}{c} n_{i} d_i cos\theta_i$, $c$ is
speed of light in vacuum, $\theta_i$ is the ray angle inside the
layer $i$ with refractive index $n_i=\sqrt{\varepsilon_i \mu_i}$,
$\eta_i=\sqrt{\varepsilon_i/\mu_i} cos\theta_i$,
$cos\theta_i=\sqrt{1-(n^2_0sin^2\theta_0/n^2_i)}$, in which $n_0$
is the refractive index of the environment wherein the incidence
wave tends to enter the structure, and $\theta_0$ is the incident
angle.

The total transfer matrix $M$ for an $N$ period structure is given
by:
\begin{eqnarray}
\left(%
\begin{array}{c}
  E_{1} \\
  H_{1} \\
\end{array}%
\right)&=&M_{B}M_{A}M_{B}M_{A}\cdot\cdot\cdot M_{B}M_{A}\left(%
\begin{array}{c}
  E_{N+1} \\
  H_{N+1} \\
\end{array}%
\right)
\nonumber\\&=&M\left(%
\begin{array}{c}
  E_{N+1} \\
  H_{N+1} \\
\end{array}%
\right)=\left(%
\begin{array}{c c}
  A &  B \\
 C &  D \\
\end{array}%
\right)
 \left(%
\begin{array}{c}
  E_{N+1} \\
  H_{N+1} \\
\end{array}%
\right),
\end{eqnarray}
where
\begin{eqnarray}
M=\left(%
\begin{array}{c c}
  A &  B \\
 C &  D \\
\end{array}%
\right),
\end{eqnarray}
with the total transfer matrix $M$, we can obtain the
transmissivity $T$, it is
\begin{eqnarray}
T=|\frac{E_{N+1}}{E_{1}}|^2=|\frac{2\eta_{0}}{A\eta_{0}+B\eta_{0}\eta_{N+1}+C+D\eta_{N+1}}|^2.
\end{eqnarray}
Where $\eta_{0}=\eta_{N+1}=\sqrt{\frac{\varepsilon_0}{\mu_0}}
\cos\theta_0$. By the Eqs. (1) and (4), we can calculate the
transmissivity of one-dimensional photonic crystal.
\begin{figure}[tbp]
\includegraphics[width=9 cm]{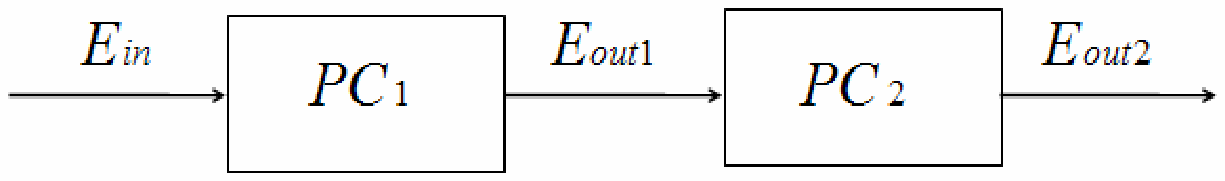}
\caption{The series connection structure one-dimensional photonic
crystal.}
\end{figure}
\begin{figure}[tbp]
\includegraphics[width=9 cm]{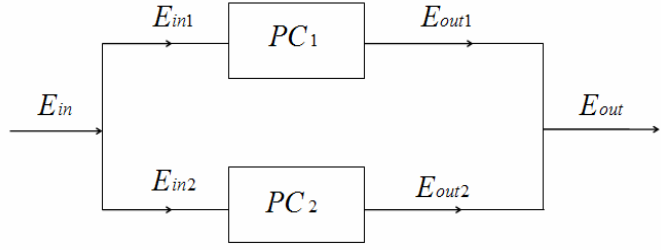}
\caption{The parallel connection structure one-dimensional
photonic crystal.}
\end{figure}
\vskip 8pt
 {\bf 3. Series connection, parallel connection and feedback one-dimensional photonic crystal transmissivity} \vskip 8pt
Two or multiple one-dimensional photonic crystal can be connected
by optical fiber, which can be designed series connection,
parallel connection and feedback compound structure
one-dimensional photonic crystal. The FIGs. 1, 2, 3 and 4 are
series connection, parallel connection, positive feedback and
negative feedback one-dimensional photonic crystal structures,
respectively. The $PC_1$ and $PC_2$ are two kinds one-dimensional
photonic crystal, $E_{in}$ is the input electric field intensity,
and $E_{out1}$ and $E_{out2}$ are the output electric field
intensity of $PC_1$ and $PC_2$.

(1) The total transmission coefficient $t$ and transmissivity $T$
of series connection one-dimensional photonic crystal are

\begin{eqnarray}
t=\frac{E_{out2}}{E_{in}}=\frac{E_{out2}}{E_{out1}}\cdot\frac{E_{out1}}{E_{in}}=t_2\cdot
t_1,
\end{eqnarray}
\begin{eqnarray}
T=|t|^2,
\end{eqnarray}
where $t_2={E_{out2}}/{E_{out1}}$ and $t_1={E_{out1}}/{E_{in}}$
for the transmission coefficients of photonic crystals $PC_1$ and
$PC_2$.

Similarly, the total transmission coefficient for $n$ photonic
crystals $PC_1$, $PC_2$, $\cdot\cdot\cdot$, $PC_n$ series
connection is
\begin{eqnarray}
t=t_n\cdot t_{n-1}\cdot\cdot\cdot t_1.
\end{eqnarray}
\begin{figure}[tbp]
\includegraphics[width=9 cm]{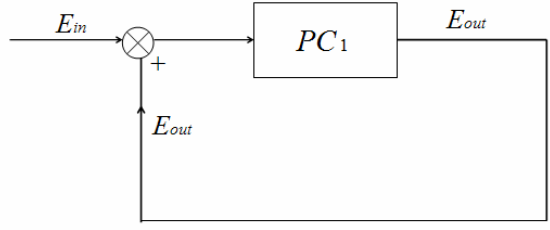}
\caption{The positive feedback structure one-dimensional photonic
crystal.}
\end{figure}

\begin{figure}[tbp]
\includegraphics[width=9 cm]{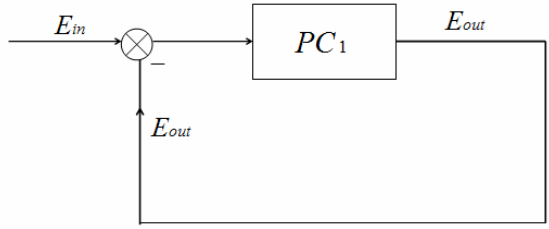}
\caption{The negative feedback structure one-dimensional photonic
crystal.}
\end{figure}
\begin{figure}[tbp]
\includegraphics[width=9 cm]{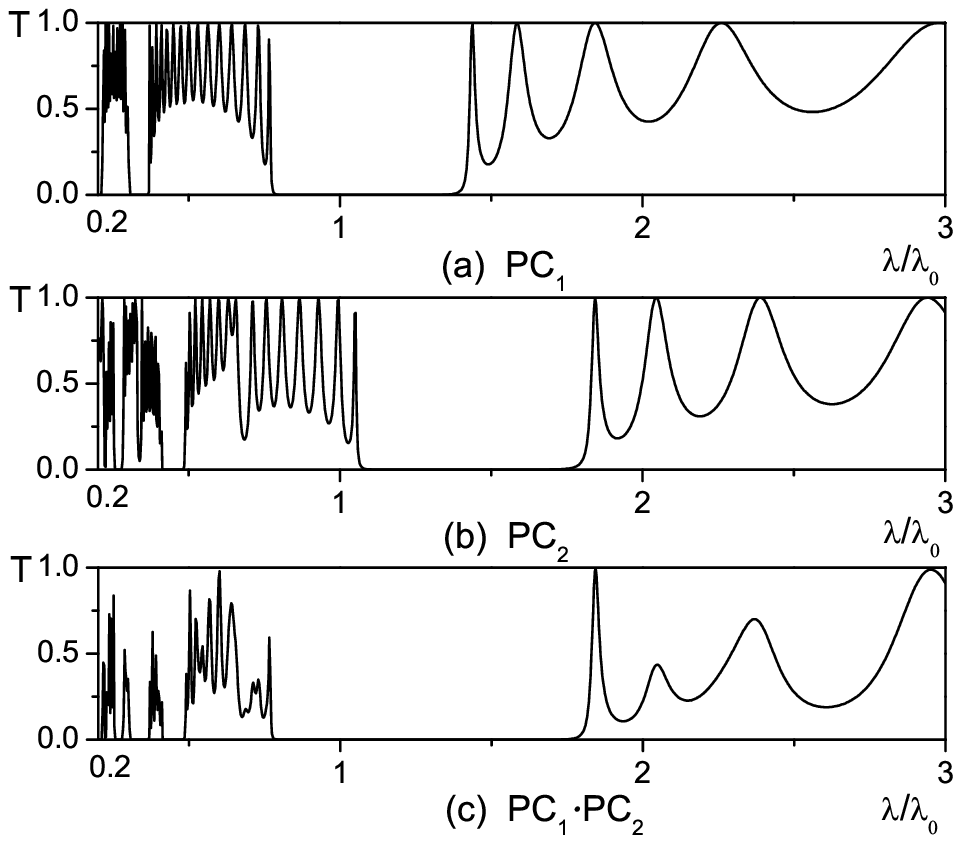}
\caption{The series connection structure transmissivity. (a)
$PC_1$ transmissivity, (b) $PC_2$ transmissivity, (c) series
connection total transmissivity.}
\end{figure}

\begin{figure}[tbp]
\includegraphics[width=9 cm]{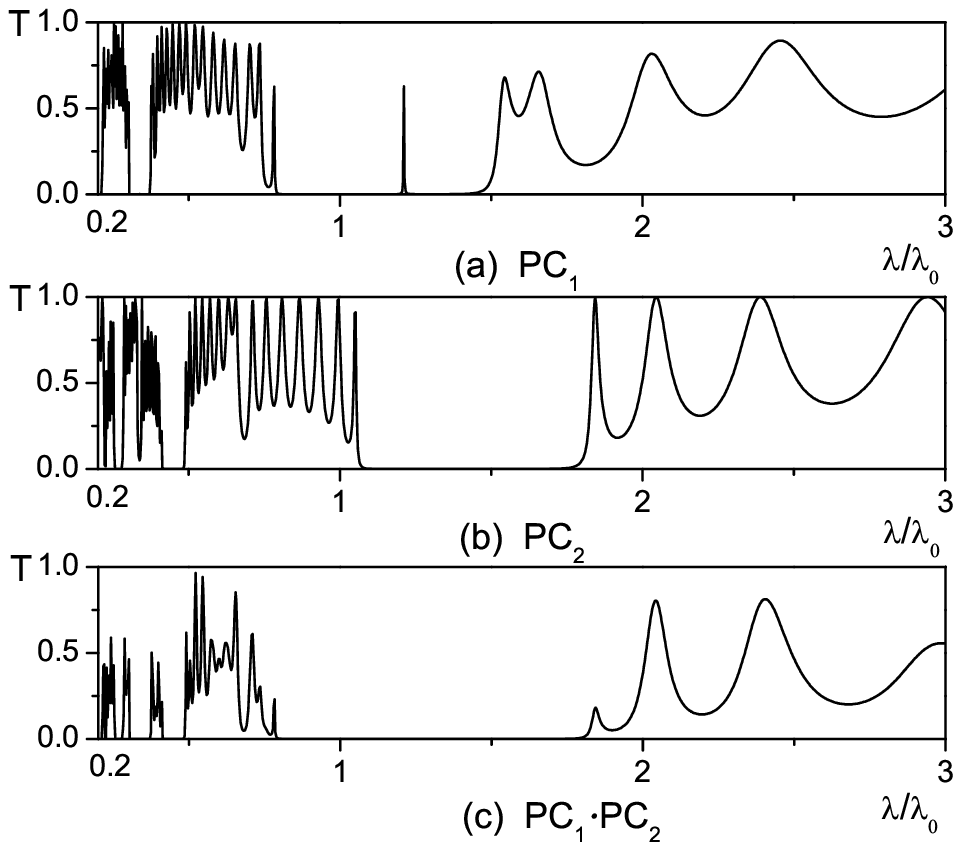}
\caption{The series connection structure transmissivity. (a)
$PC_1$ with defect layer transmissivity, (b) $PC_2$
transmissivity, (c) series connection total transmissivity.}
\end{figure}
(2) The total transmission coefficient $t$ and transmissivity $T$
of parallel connection one-dimensional photonic crystal are
\begin{eqnarray}
t_{\pm}=\frac{E_{out}}{E_{in}}=\frac{E_{out1}\pm
E_{out2}}{E_{in}}=\frac{t_1E_{in1}\pm t_2E_{in2}}{E_{in}},
\end{eqnarray}
when $E_{out1}$ and $E_{out2}$ phase are the same (opposite), the
numerator of Eq. (8) takes $+$ ($-$).

when $t_1=t_2=t$, there are
\begin{eqnarray}
t_{+}=t_1=t_2=t, \hspace{0.5in} t_{-}=\frac{t(E_{in1}-
E_{in2})}{E_{in}}=t(c_1-c_2),
\end{eqnarray}
where $c_1={E_{in1}}/{E_{in}}$ and $c_2={E_{in2}}/{E_{in}}$.

when $t_1\neq t_2$, there is
\begin{eqnarray}
t_{+}=c_1t_1+ c_2t_2, \hspace{0.5in} t_{-}=c_1t_1- c_2t_2
\hspace{0.1in} (c_1+c_2=1),
\end{eqnarray}
Similarly, the total transmission coefficient of $n$ photonic
crystals $PC_1$, $PC_2$, $\cdot\cdot\cdot$, $PC_n$ parallel
connection is
\begin{eqnarray}
t_{\pm}&=&\frac{E_{out}}{E_{in}}=\frac{E_{out1}\pm E_{out2}\pm
\cdot\cdot\cdot \pm
E_{outn}}{E_{in}}\nonumber\\
&=&\frac{t_1E_{in1}\pm t_2E_{in2}\pm
\cdot\cdot\cdot \pm t_nE_{inn}}{E_{in}}\nonumber\\
&=&c_1t_1\pm c_2t_2\pm \cdot\cdot\cdot \pm c_nt_n,
\end{eqnarray}
(3) We can calculate the total transmission coefficient $t$ and
transmissivity $T$ of positive feedback one-dimensional photonic
crystal.

The output of first times positive feedback is
\begin{eqnarray}
E_{out1}=tE_{in},
\end{eqnarray}
the output of second times positive feedback is
\begin{eqnarray}
E_{out2}=t(E_{in}+E_{out1})=t(t+1)E_{in},
\end{eqnarray}
the transmission coefficient of second times positive feedback is
\begin{eqnarray}
t_2=\frac{E_{out2}}{E_{in}}=t^2+t,
\end{eqnarray}
the output of third times positive feedback is
\begin{eqnarray}
E_{out3}=t(E_{in}+E_{out2})=t(E_{in}+t(t+1)E_{in})=(t^3+t^2+t)E_{in},
\end{eqnarray}
the transmission coefficient of third times positive feedback is
\begin{eqnarray}
t_3=\frac{E_{out3}}{E_{in}}=t^3+t^2+t,
\end{eqnarray}
similarly, the transmission coefficient of $n-th$ times positive
feedback is
\begin{eqnarray}
t_n=\frac{E_{outn}}{E_{in}}=t^n+t^{n-1}+\cdot\cdot\cdot +t),
\end{eqnarray}
finally, we can obtain the limit transmission coefficient of
positive feedback, it is
\begin{eqnarray}
t_\infty=\lim_{n\rightarrow \infty} t_n=\lim_{n\rightarrow \infty}
(t^n+t^{n-1}+\cdot\cdot\cdot +t)=\frac{t}{1-t},
\end{eqnarray}

(4) We can calculate the total transmission coefficient $t$ and
transmissivity $T$ of negative feedback one-dimensional photonic
crystal.

The output of first times negative feedback is
\begin{eqnarray}
E_{out1}=tE_{in},
\end{eqnarray}
the output of second times negative feedback is
\begin{eqnarray}
E_{out2}=t(E_{in}-E_{out1})=t(1-t)E_{in},
\end{eqnarray}
the transmission coefficient of second times negative feedback is
\begin{eqnarray}
t_2=\frac{E_{out2}}{E_{in}}=t-t^2,
\end{eqnarray}
the output of third times negative feedback is
\begin{eqnarray}
E_{out3}=t(E_{in}-E_{out2})=t(E_{in}-t(1-t)E_{in})=(t-t^2+t^3)E_{in},
\end{eqnarray}
the transmission coefficient of third times negative feedback is
\begin{eqnarray}
t_3=\frac{E_{out3}}{E_{in}}=t-t^2+t^3,
\end{eqnarray}
similarly, the transmission coefficient of $n-th$ times negative
feedback is
\begin{eqnarray}
t_n=\frac{E_{outn}}{E_{in}}=t-t^2+t^3-t^4+\cdot\cdot\cdot
(-1)^{n+1}t^n,
\end{eqnarray}
finally, we can obtain the limit transmission coefficient of
negative feedback, it is
\begin{eqnarray}
t_\infty=\lim_{n\rightarrow \infty} t_n=\lim_{n\rightarrow \infty}
(t-t^2+t^3-t^4+\cdot\cdot\cdot (-1)^{n+1}t^n)=\frac{t}{1+t}.
\end{eqnarray}

\begin{figure}[tbp]
\includegraphics[width=9 cm]{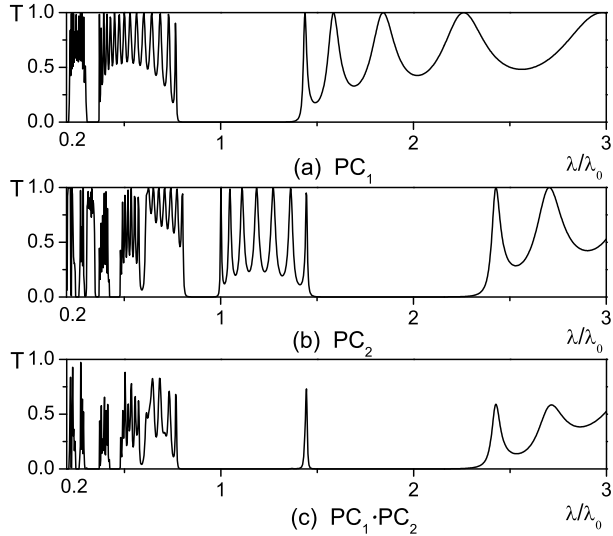}
\caption{The series connection structure transmissivity. (a)
$PC_1$ transmissivity, (b) $PC_2$ transmissivity, (c) series
connection total transmissivity including model.}
\end{figure}

\begin{figure}[tbp]
\includegraphics[width=9 cm]{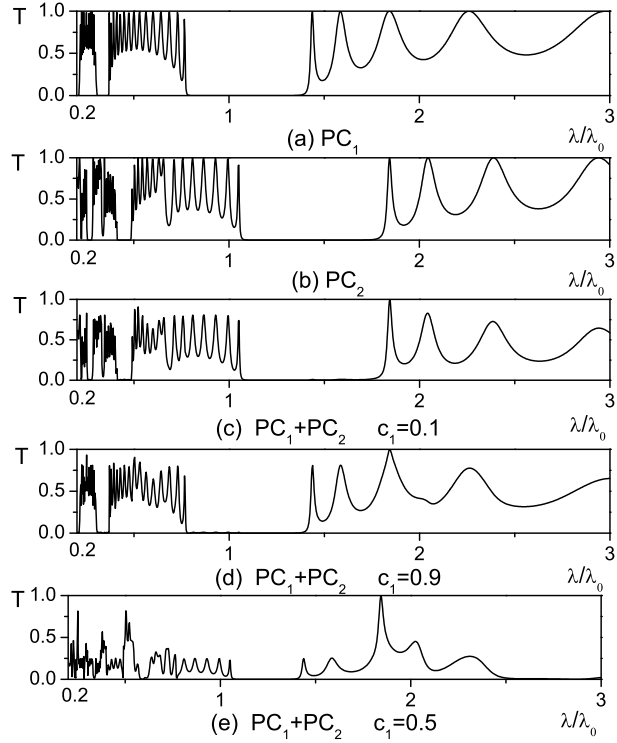}
\caption{The parallel connection structure transmissivity. (a)
$PC_1$ transmissivity, (b) $PC_2$ transmissivity, (c) parallel
connection total transmissivity $c_1=0.1$, (d) parallel connection
total transmissivity $c_1=0.9$, (e) parallel connection total
transmissivity $c_1=0.5$.}
\end{figure}

\begin{figure}[tbp]
\includegraphics[width=9 cm]{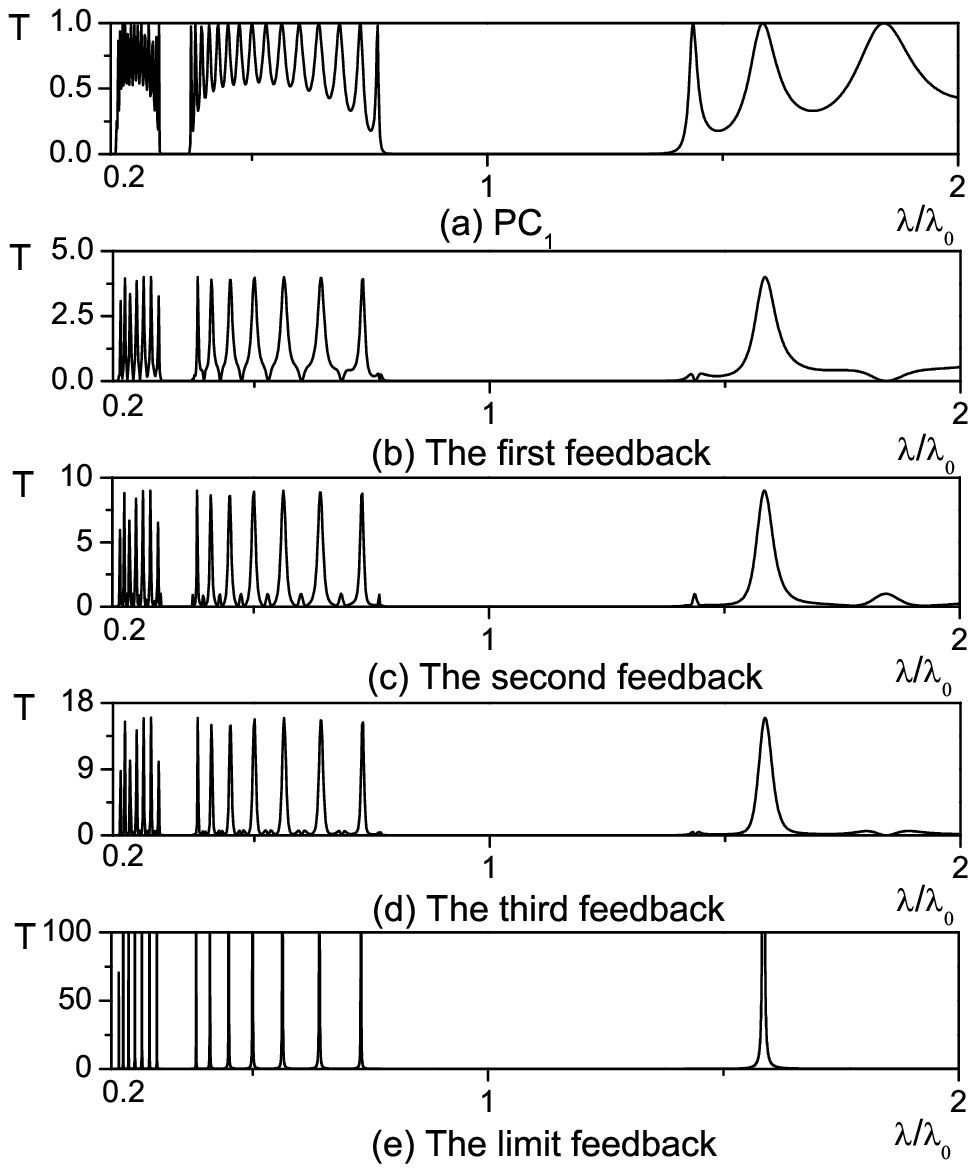}
\caption{The positive feedback structure transmissivity. (a)
$PC_1$ transmissivity, (b) the first feedback transmissivity, (c)
the second feedback transmissivity, (d) the third feedback
transmissivity, (e) the limit feedback transmissivity.}
\end{figure}

\begin{figure}[tbp]
\includegraphics[width=9 cm]{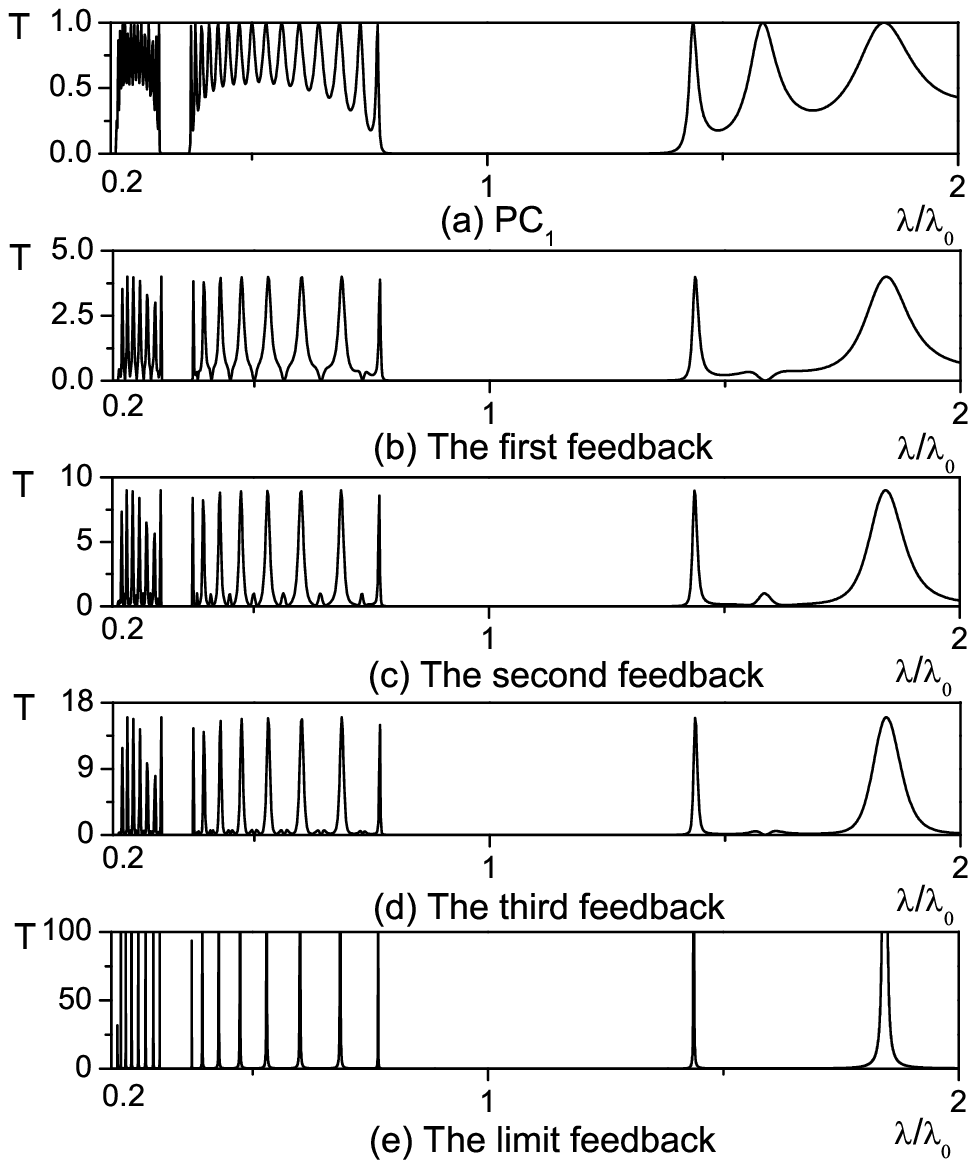}
\caption{The negative feedback structure transmissivity. (a)
$PC_1$ transmissivity, (b) the first feedback transmissivity, (c)
the second feedback transmissivity, (d) the third feedback
transmissivity, (e) the limit feedback transmissivity.}
\end{figure}

\begin{figure}[tbp]
\includegraphics[width=9 cm]{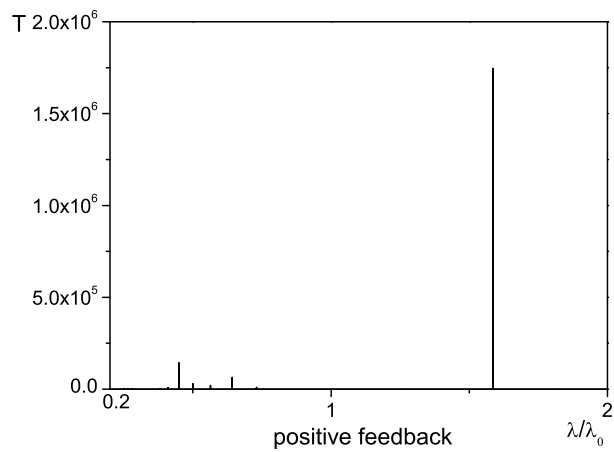}
\caption{The transmissivity maximum value of positive feedback
structure.}
\end{figure}

\begin{figure}[tbp]
\includegraphics[width=9 cm]{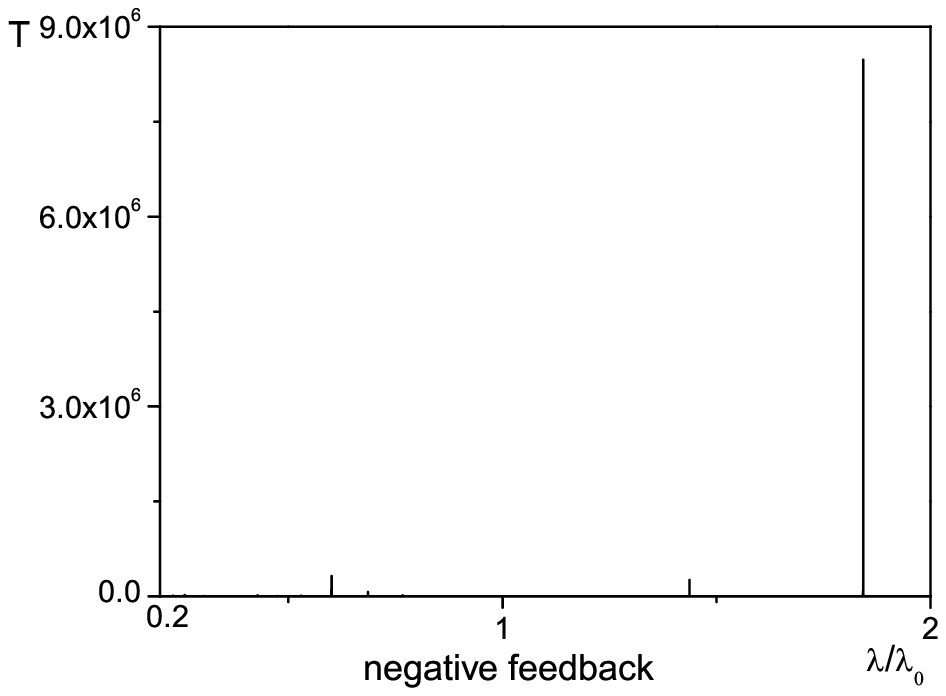}
\caption{The transmissivity maximum value of negative feedback
structure.}
\end{figure}
 \vskip 8pt
 {\bf 4. Numerical result}
 \vskip 8pt

In this section, we report our numerical results of compound
structure one-dimensional photonic crystal, the two kinds of
photonic crystal $PC_1$ and $PC_2$, their structures are
$(AB)^{8}$ and $(CD)^{8}$, respectively. The main parameters are:
The center angle wavelength $\lambda_0=1.55\cdot10^{-6}m$, the
medium $A$ refractive indices $n_a=1.45$, thickness $a=267nm$, the
medium $B$ refractive indices $n_b=3.59$, thickness $b=108nm$, the
medium $C$ refractive indices $n_c=1.76$, thickness $c=306nm$, the
medium $D$ refractive indices $n_d=3.89$, thickness $d=128nm$.
Firstly, we study the series connection structure photonic
crystal, which is constituted by $PC_1$ and $PC_2$, it is shown in
FIG. 1. The series connection structure is referred to as
$PC_1\cdot PC_2$. From Eqs. (5) and (6), we can calculate the
series connection structure transmissivity. The FIG. 5 (a), (b)
and (c) are the transmissivity corresponding to the structure
$PC_1$, $PC_2$ and $PC_1\cdot PC_2$. From FIG. 5 (a), (b) and (c),
we can obtain some results: (1) The forbidden band width of series
connection photonic crystal becomes more wider, and it is the
union of corresponding forbidden band of part photonic crystal
$PC_1$ and $PC_2$, which is similar as the series connection ohm
law in circuit. (2) We can obtain the more wider forbidden band by
photonic crystal series connection. (3) With the number of series
connection photonic crystal increasing, the total width of
forbidden band increase. In FIG. 6, the photonic crystal $PC_1$
with defect layer, its structure is $(AB)^{4}F(AB)^{4}$, the
defect medium $F$ parameters are: refractive indices $n_f=2.58$,
thickness $f=256nm$. There is a defect model in FIG. 6 (a). When
the $PC_1$ and $PC_2$ constitute series connection structure
photonic crystal $PC_1\cdot PC_2$, its forbidden band width is the
union of corresponding forbidden band of $PC_1$ and $PC_2$, and
the defect model disappears, since the defect model is in the
forbidden band range of $PC_1$ and $PC_2$. In FIG. 7, we change
the thickness $c$ of $PC_2$, it is $c=506nm$, the position of
forbidden band red-shift. When the $PC_1$ and $PC_2$ constitute
series connection structure photonic crystal $PC_1\cdot PC_2$, its
forbidden band width is the union of corresponding forbidden band
of $PC_1$ and $PC_2$, the forbidden band width increases
obviously. There is a defect model in series connection photonic
crystal $PC_1\cdot PC_2$ forbidden band, it is because there is a
little conduction band overlap of $PC_1$ and $PC_2$ in $PC_1\cdot
PC_2$ forbidden band. Secondly, we study the parallel connection
structure photonic crystal, which is constituted by $PC_1$ and
$PC_2$, it is shown in FIG. 2. The parallel connection structure
is referred to as $PC_1+PC_2$. By Eq. (10), we can calculate the
parallel connection structure transmissivity $|t_+|^2$. The FIG. 8
(a) and (b) are the transmissivity of $PC_1$ and $PC_2$, and (c),
(d) and (e) are the transmissivity of parallel connection
$PC_1+PC_2$ corresponding coefficient $c_1$ is $0.1$, $0.9$ and
$0.5$, respectively. We can obtain some results: (1) When
coefficient $c_1=0.1$ the transmissivity of parallel connection
(FIG. 8 (c)) is close to the transmissivity of $PC_2$ (FIG. 8
(b)). (2) When coefficient $c_1=0.9$ the transmissivity of
parallel connection (FIG. 8 (d)) is close to the transmissivity of
$PC_1$ (FIG. 8 (a)). (3) When coefficient $c_1=0.5$ the forbidden
band of parallel connection (FIG. 8 (e)) is the intersection of
corresponding forbidden band of $PC_1$ and $PC_2$. Finally, we
study the positive and negative feedback structure photonic
crystal, which is constituted by $PC_1$, it is shown in FIGs. 3
and 4. The positive and negative feedback structure is referred to
as $PC_1+$ and $PC_1-$, respectively. From Eqs. (12) to (18), we
can calculate the positive feedback structure transmissivity. The
FIG. 9 (a) is the transmissivity of $PC_1$, (b), (c), (d) and (e)
are the transmissivity of the first time, the second time, the
third time and the limit positive feedback, respectively. We can
obtain some results: (1) There are the same forbidden band width
for the first time, the second time, the third time and the limit
positive feedback. (2) From the first times positive feedback, the
transmissivity is larger than $1$, which can achieve light
amplification in positive feedback. (3) With the increasing of
feedback times, the amplitude of transmissivity increases. From
Eqs. (19) to (25), we can calculate the negative feedback
structure transmissivity. The FIG. 10 (a) is the transmissivity of
$PC_1$, (b), (c), (d) and (e) are the transmissivity of the first
time, the second time, the third time and the limit negative
feedback, respectively. We can obtain some results: (1) There are
the same forbidden band width for the first time, the second time,
the third time and the limit negative feedback. (2) From the first
times negative feedback, the transmissivity is larger than $1$,
which can achieve light amplification in negative feedback. (3)
With the increasing of feedback times, the amplitude of
transmissivity increases. The FIGs. 11 and 12 are the positive and
negative limit feedback maximum transmissivity. We can obtain some
results: (1) At $\lambda/\lambda_0=1.5851$, the positive limit
feedback maximum transmissivity $T_m=1.7458\cdot10^6$. At
$\lambda/\lambda_0=1.8427$, the negative limit feedback maximum
transmissivity $T_m=8.4808\cdot10^6$, there are the extremely
strong output light intensity, which can be designed into the
photonic crystal laser. (2) The positive and negative feedback
photonic crystal both can achieve the light amplification, which
is different from the electronic positive and negative feedback
system, only the electronic positive system can achieve the
electrical signal amplification.

\vskip 8pt {\bf 5. Conclusion}
 \vskip 8pt

In summary, we have proposed a new compound structure
one-dimensional photonic crystal, which include series connection,
parallel connection and positive and negative feedback compound
structure photonic crystal. We have calculated their transmission
characteristics and obtained some new results: (1) The forbidden
band width of series connection photonic crystal becomes more
wider, and it is the union of corresponding forbidden band of part
photonic crystal, which is similar as the series connection ohm
law in circuit. (2) We can obtain the more wider forbidden band by
photonic crystal series connection. (3) With the number of series
connection photonic crystal increasing, the total width of
forbidden band increase. (4) The total forbidden band width of
parallel connection photonic crystal is adjusted by superposition
coefficient. (5) For the positive and negative feedback, the
transmissivity are larger than $1$, which can achieve light
amplification in the positive and negative feedback photonic
crystal. (6) At $\lambda/\lambda_0=1.5851$, the positive limit
feedback maximum transmissivity $T_m=1.7458\cdot10^6$. At
$\lambda/\lambda_0=1.8427$, the negative limit feedback maximum
transmissivity $T_m=8.4808\cdot10^6$, there are the extremely
strong output light intensity, which can be designed into the
photonic crystal laser. All these results should be help to design
new type optical devices, such as optical amplifier, photonic
crystal laser and so on.

\vskip 8pt {\bf 6.  Acknowledgment} \vskip 8pt

This work is supported by Scientific and Technological Development
Foundation of Jilin Province, Grant Number: 20130101031JC.

\end{document}